\documentclass[a4paper,11pt]{article}
\pdfoutput=1 
\usepackage[utf8]{inputenc}
\usepackage{jheppub} 
\usepackage[T1]{fontenc} 
\graphicspath{{img/}} 

\title{\boldmath Multi-charged moments and symmetry-resolved R\'enyi entropy of free compact boson for multiple disjoint intervals}

 \author[a]{Himanshu Gaur}
 \author[a,b]{and Urjit A. Yajnik,}
 \affiliation[a]{Department of Physics, Indian Institute of Technology Bombay, Powai, Mumbai, Maharashtra 400076 India}
\affiliation[b]{Department of Physics, Indian Institute of Technology Gandhinagar, Gujarat, 382055 India}

\emailAdd{194123018@iitb.ac.in}
\emailAdd{yajnik@iitb.ac.in}

\abstract{We study multi-charged moments and symmetry-resolved R\'enyi entropy of free compact boson for multiple disjoint intervals. The R\'enyi entropy evaluation involves computing the partition function of the theory on Riemann surfaces with genus $g>1$. This makes R\'enyi entropy sensitive to the local conformal algebra of the theory. The free compact boson possesses a global $U(1)$ symmetry with respect to which we resolve R\'enyi entropy. The multi-charged moments are obtained by studying the correlation function of flux-generating vertex operators on the associated Riemann surface. Symmetry-resolved R\'enyi entropy is then obtained from the Fourier transforms of the charged moments. R\'enyi entropy is shown to have the familiar equipartition into local charge sectors upto the leading order. The multi-charged moments are also essential in studying the symmetry resolution of mutual information. The multi-charged moments of the self-dual compact boson and massless Dirac fermion are also shown to match for the cases when the associated reduced density matrix moments are known to be the same. Finally, we numerically check our results against the tight-binding model.}

\makeatletter
\gdef\@fpheader{}
\makeatother
\begin{document} 
\maketitle
\flushbottom

\section{Introduction} \label{introduction}
The phenomenon of entanglement has proved to be a cornerstone of quantum theories. Entanglement has been at the centre of the recent development in many areas of physics, especially in quantum computation \cite{nielsen2010quantum}, critical quantum many-body systems \cite{amico2008entanglement}, gauge-gravity duality \cite{ryu2006holographic, ryu2006aspects} and black-hole entropy \cite{solodukhin2011entanglement}.  In $1$d critical systems, entanglement for a single interval is sensitive to the central charge of the corresponding conformal field theory \cite{vidal2003entanglement}. To study entanglement in the state $|{\Psi}\rangle$ of a quantum system, the system is first partitioned into two subsystems, subsystem $A$ and its complement $B$, such that the Hilbert space is $\mathcal{H}=\mathcal{H}_A\otimes\mathcal{H}_B$. Among several measures of entanglement, the R\'enyi entropies are the most prominent. R\'enyi entropies $S_n$ are given by
\begin{equation} \label{eq1.1}
S_n=\frac{1}{1-n}\ln\mathrm{Tr}\rho_A^n,
\end{equation}
where $n$ is a positive integer, and the reduced density matrix $\rho_A$ is given by $\mathrm{Tr}_B|{\Psi}\rangle\langle{\Psi}|$. The $n\to 1$ limit of eq.\eqref{eq1.1} is just the entanglement entropy. The R\'enyi entropies for a $2$d conformal field theories with a central charge $c$ are found to be proportional to the central charge $c$ and scales as the logarithm of length of the subsystem $A$ \cite{holzhey1994geometric,calabrese2004entanglement,calabrese2009entanglement}.

We now consider the scenario when the subsystem $A$ is composed of multiple disjoint intervals $A_i$, $A=\cup_{i}A_i$. Entanglement in $2$d conformal field theories in this scenario now also becomes sensitive to the local conformal operator content as well \cite{calabrese2009entanglement2,coser2014renyi}. The evaluation of R\'enyi entropies in field theories involves computing the partition function on the replica surface. In the present case, the replica surface is a Riemann surface with a genus $g>0$. The partition function of modular invariant conformal field theories on such surfaces shows interesting behaviour and thus so does entanglement. Entanglement measures have been substantially investigated for disjoint intervals in critical systems \cite{furukawa2009mutual,calabrese2009entanglement2,calabrese2011entanglement,rajabpour2012entanglement, coser2014renyi,alba2011entanglement,coser2016spin,fagotti2010entanglement,calabrese2012entanglement, coser2016towards, headrick2013bose, casini2005entanglement}.

Entanglement studies for quantum systems with global internal symmetries can be made more refined. The entanglement measures of such systems for certain states decompose into the local charge sectors of the subsystem $A$ corresponding to the global symmetry. The studies which resolve entanglement into the local charge sectors have been termed symmetry-resolved entanglement. Since the seminal work of \cite{goldstein2018symmetry}, similar studies on symmetry-resolved entanglement have been made in different contexts. In critical systems, symmetry resolution of entanglement entropies \cite{xavier2018equipartition, turkeshi2020entanglement, bonsignori2019symmetry,  fraenkel2020symmetry,ares2022symmetry, jones2022symmetry,2023,barghathi2019operationally, barghathi2018renyi,ghasemi2023universal, murciano2020entanglement, murciano2020symmetry, murciano2020symmetry1, ares2022multi, foligno2023entanglement, capizzi2022entanglement, horvath2021u, capizzi2022renyi, capizzi2022renyi2, capizzi2023full, parez2021exact, parez2021quasiparticle, estienne2021finite}, relative entropies \cite{capizzi2021symmetry, chen2021symmetry}, operator entanglement \cite{rath2023entanglement, murciano2023more} and negativity \cite{cornfeld2018imbalance,murciano2021symmetry, gaur2023charge, chen2022charged, chen2022dynamics,chen2023dynamics, feldman2019dynamics, parez2022dynamics, berthiere2023reflected} have been studied largely for $U(1)$ symmetry. Symmetry resolved entanglement of Wess-Zumino-Witten models have also been studied in \cite{calabrese2021symmetry}. Symmetry resolved entanglement of $2$d CFTs have also been shown to have an interesting relation with the boundary conformal field theory \cite{di2023boundary, kusuki2023symmetry, PhysRevLett.131.151601}. Similar studies have been made in the context of gravity as well \cite{zhao2021symmetry, weisenberger2021symmetry, zhao2022charged, belin2013holographic, milekhin2021charge, gaur2023symmetry}. The protocols for measuring symmetry resolved entanglement have been discussed in \cite{neven2021symmetry,cornfeld2018imbalance}. 

In the present work, we study the symmetry-resolved R\'enyi entropies of free compact boson for an arbitrary number of disjoint intervals. We also evaluate the multi-charged moments of the free compact boson in the same settings. The multi-charged moments obtained here are essential for studying symmetry-resolved mutual information. Mutual information measures quantify the entanglement between the intervals themselves, however, they are not a pure measure of correlations only and hence can be negative sometimes \cite{kormos2017temperature}.

\begin{figure}
\centering 
\includegraphics[width=0.8\textwidth]{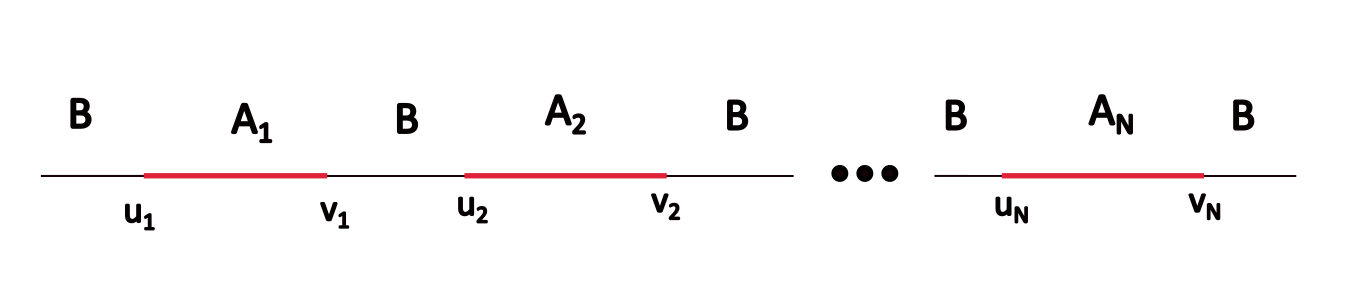}
\caption{\label{fig:i} Multiple disjoint intervals. The interval $A_i$ is  $(u_1,v_1)$, where $v_i>u_i$ and its length is denoted  $\ell_1=|v_1-u_1|$.}
\end{figure}

To set the stage for calculations later, we consider $N$ disjoint intervals $A_i$ whose boundary points are denoted $u_i$ and $v_i$, see Figure \ref{fig:i}. The length of each interval $A_i$ is denoted $\ell_i$.

The organisation of this work is as follows. In Section \ref{section2} we discuss the symmetry-resolved R\'enyi entropy, and the multi-charged moments of theories with a global $U(1)$ symmetry and also discuss the replica method for multi-charged moments in CFTs. In Section \ref{section3}, we discuss the theory of free compact boson and review the R\'enyi entropy results in the theory. In Section \ref{section4}, we evaluate the charged and multi-charged moments of free compact boson for multiple disjoint intervals. In Section \ref{section5}, we evaluate the symmetry resolved R\'enyi entropy of free compact boson in the same settings. Finally, in Section \ref{section6} we give a conclusion of the present work. We also have four appendices. In Appendices \ref{A}, \ref{B}, and \ref{D} some necessary computations and calculations are discussed. In Appendix \ref{C} we evaluate the multi-charged moments of massless Dirac fermion for multiple disjoint intervals. 
\section{Symmetry resolution of Entanglement} \label{section2}
In this section we discuss the symmetry resolution of R\'enyi entropy for theories with a global $U(1)$ symmetry \cite{goldstein2018symmetry}. We also discuss the charged and the multi-charged moments in the same context. Finally, we revisit the replica trick for evaluating the multi-charged moments in CFTs.
\subsection{Symmetry Resolved R\'enyi Entropy}
We consider a theory with a global internal $U(1)$ symmetry and consider its bipartition into subsystem $A$ and its complement $B$. Let the theory be in a state that satisfies $[\rho,\hat{Q}]=0$, where $\rho$ is the density matrix and $\hat{Q}$ is the charge operator corresponding to the $U(1)$ symmetry. Then for subsystem $A$ we have $[\rho_{A},\hat{Q}_{A}]=0$, where $\rho_A$ is the reduced density matrix and $\hat{Q}_A$ is the local charge operator in $A$. This implies that $\rho_A$ is block diagonal in local charge sectors $q$, here $q$ are the eigenvalues of $\hat{Q}_{A}$. This allows us to study entanglement in local charge sectors $q$, such studies have been termed symmetry-resolved entanglement.

The reduced density matrix in the charge sector $q$, $\rho_{A,q}$ is defined as
\begin{equation} \label{eq2.1}
p_q\rho_{A,q}=\Pi_q\rho_A\Pi_q,
\end{equation} 
where $\Pi_q$ is the projection operator on to the charge sector $q$ and $p_q$ is the probability of the subsystem $A$ having the charge $q$, mathematically $p_q=\mathrm{Tr}[\rho_{A}\Pi_q]$. The R\'enyi entropies in these charge sectors are then defined as 
\begin{equation}
S_{n,q}=\frac{1}{1-n}\ln\left[\mathrm{Tr}\rho_{A,q}^n\right]. \label{eq2.2}
\end{equation}
The $n\to 1$ limit of \eqref{eq2.2} is the symmetry-resolved entanglement entropy. However, $\Pi_q$ may not always be readily constructed, as is the case for our present interests. In such cases charge moments $Z_n(\alpha)$ are introduced
\begin{equation}\label{eq2.3}
Z_{n}(\alpha)=\mathrm{Tr}\left[\rho_A^n e^{i\alpha\hat{Q}_A}\right].
\end{equation}
Since, in the present case of our interest $A=\bigcup_{i=1}^{N}A_i$, it is also useful to define the more generalised quantities $Z_{N,n}(\alpha_1,\cdots,\alpha_N)$ \cite{parez2021exact} as
\begin{equation} \label{eq2.4}
Z_{N,n}(\alpha_1,\cdots,\alpha_N)=\mathrm{Tr}\left[\rho_A^n e^{i\alpha_1\hat{Q}_{A_1}+\cdots+i\alpha_N\hat{Q}_{A_N}}\right].
\end{equation}
The operator $\hat{Q}_{A_i}$ in \eqref{eq2.4} is the local charge operator corresponding to the interval $A_i$. The quantities $Z_{N,n}(\alpha_1,\cdots,\alpha_N)$ have been termed multi-charged moments. Their utility will become clear in a moment. We denote by $\mathcal{Z}_{N,n}(q_1,\cdots,q_N)$, the Fourier transforms of the multi-charged moments 
\begin{equation} \label{2.5}
\mathcal{Z}_{N,n}(q_1,\cdots,q_N)=\prod_{j=1}^{N}\left(\frac{1}{2\pi}\int_{-\pi}^{\pi}\mathrm{d}\alpha_{j}e^{-iq_{j}\alpha_{j}}\right)Z_{N,n}(\alpha_1,\cdots,\alpha_N)
\end{equation} 
The quantity $\mathcal{Z}_{N,1}(q_1,\cdots,q_N)$ is the joint probability of finding the charges $q_j$ in the subregions $A_j$ respectively. Similarly using \eqref{eq2.3}, we also define the quantities $\mathcal{Z}_{N,n}(\alpha)$ as
\begin{equation} \label{2.6}
\mathcal{Z}_{N,n}(q)=\frac{1}{2\pi}\int_{-\pi}^{\pi}\mathrm{d}\alpha e^{-iq\alpha}Z_{N,n}(\alpha).
\end{equation}
The symmetry resolved R\'enyi entropy using $\mathcal{Z}_n(\alpha)$ is given by
\begin{equation}
S_{N,n,q}=\frac{1}{1-n}\ln\left[\frac{\mathcal{Z}_{N,n}(q)}{\mathcal{Z}_{N,1}^n(q)}\right]. \label{eq2.7}
\end{equation}
We remark that the multi-charged moments given by \eqref{eq2.4} are also essential in the evaluation of symmetry-resolved mutual information, however, we do not consider symmetry-resolved mutual information in the present work.
\subsection{Replica trick}
We again consider the bipartite $A$, and its complement $B$ and further assume that the theory is in its ground state. The $n^{th}$ moment of the reduced density matrix, $\mathrm{Tr}\rho_{A}^n$ is proportional to the partition function $Z_n$ of the theory on the Replica surface $\Sigma_{n}$,
\begin{equation} \label{eq2.8}
\mathrm{Tr}\rho_{A}^n=Z_{1}^{-n}\int\mathrm{D}\phi e^{-\int_{\Sigma_{n}}\mathrm{d}^{n}x\mathcal{L}_{E}[\phi]},
\end{equation}
where $Z_{1}$ is the partition function on the plane and $\mathcal{L}_E$ is the euclidean Lagrangian of the theory. The surface $\Sigma_n$ is generally a Riemann surface. In our case, we are dealing with a $2\mathrm{d}$ theory with the decomposition $A=\bigcup_{i=1}^{N}A_i$. The corresponding Riemann surface has a genus $(N-1)\times(n-1)$.
\begin{figure}
\centering 
\includegraphics[width=0.8\textwidth]{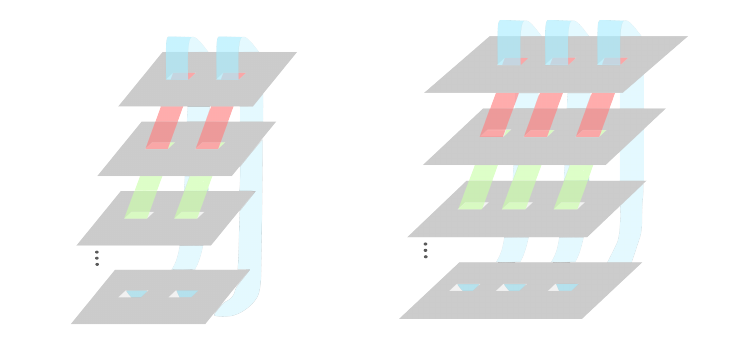}
\caption{\label{fig:ii} Riemann Surface $\Sigma_{N,n}$ obtained by sewing together $n$ copies of $\mathrm{\rho^n}$ in the evaluation of $n^{th}$ moment of the reduced matrix in the case of two intervals and three intervals respectively.}
\end{figure}

When dealing with $2\mathrm{d}$ CFT's it is particularly helpful to consider $n$ copies of the theory $\phi$ on the plane denoted $\{\phi_{i}\}$ and introduce the twist field $\mathcal{T}_n$ \cite{calabrese2004entanglement,cardy2008form}. These twist fields correspond to the cyclic symmetry:
\begin{equation} \label{eq2.9}
\begin{split}
\mathcal{T}_n\: :\: i\to i+1\: \mod n, \\
\mathcal{\bar{T}}_n\: :\: i\to i-1\: \mod n.
\end{split}
\end{equation}
In this framework, the partition function $Z_n$ is given by the correlation function of the twist fields,
\begin{equation} \label{eq2.10}
Z_n\propto \left\langle\prod_{i=1}^{N}\mathcal{T}_n(u_i)\mathcal{\bar{T}}_n(v_i)\right\rangle.
\end{equation}
The twist fields $\mathcal{T}_n$ and $\mathcal{\bar{T}}_n$ are primary conformal fields with the scaling dimension \cite{calabrese2004entanglement}
\begin{equation} \label{eq2.11}
\Delta_n=\frac{c}{12}\left(\frac{1-n^2}{n}\right),
\end{equation}
where $c$ is the central charge of the theory.

The presence of factors $e^{i\alpha_i\hat{Q}_{A_{i}}}$ in eq.\eqref{eq2.4} modifies the boundary conditions between the different sheets on the Riemann surface $\Sigma_{N,n}$. The multi-charged moments may be computed by introducing the flux-generating vertex operators $\mathcal{V}_{\alpha_i}$ \cite{goldstein2018symmetry}. The operators $\mathcal{V}_{\alpha_i}$ generate the desired boundary conditions on the Riemann surface. The vertex operators $\mathcal{V}_{\alpha_i}$ and $\mathcal{V}_{-\alpha_i}$ are placed on the boundary points $u_i$ and $v_i$ of the interval $A_{i}$ respectively, on the Riemann surface $\Sigma_{N,n}$. The multi-charged moments are proportional to the correlation function
\begin{equation} \label{eq2.12}
Z_{N,n}(\alpha_1,\cdots,\alpha_N)\propto \left\langle\prod_{i=1}^{N}\mathcal{V}_{\alpha_i}(u_i)\mathcal{V}_{-\alpha_i}(v_i)\right\rangle_{\Sigma_{N,n}}Z_{N,n},
\end{equation}
on the Riemann surface $\Sigma_{N,n}$, where $Z_{N,n}$ in the partition function on $\Sigma_{N,n}$. Similarly the charged moments $Z_n(\alpha)$ are given by
\begin{equation} \label{eq2.13}
Z_{N,n}(\alpha)\propto \left\langle\prod_{i=1}^{N}\mathcal{V}_{\alpha}(u_i)\mathcal{V}_{-\alpha}(v_i)\right\rangle_{\Sigma_{N,n}}Z_{N,n}.
\end{equation}
\section{Free Compact Boson} \label{section3}
In this section, we discuss the theory of $2\mathrm{d}$ free compact boson. We also revisit the results for the R\'enyi entropy of free compact boson when $N\geq 2$.

Free compact boson in $2\mathrm{d}$ is a conformally invariant theory with central charge $c=1$. It is also the continuum theory of Luttinger liquid. The theory is described by the Lagrangian
\begin{equation} \label{eq3.1}
\mathcal{S}=\frac{1}{8\pi}\int\mathrm{d}^2x\,\partial_{\mu}\varphi\partial^{\mu}\varphi.
\end{equation}
The target space of the theory is a circle of circumference $2\pi R$, where the compactification radius $R$ is related to the Luttinger parameter $K$ by $K=\sqrt{\frac{2}{R}}$. The theory has a duality under $R \to {2}/{R}$ and is self-dual at $R=\sqrt{2}$. This duality is known as the T-duality. 

Free compact boson possesses $U(1)\times U(1)$ symmetry. The theory is invariant under the translations in the target space $\varphi\to\varphi+a$, this corresponds to a $U(1)$ symmetry, sometimes denoted $U_{shift}(1)$. The second $U(1)$ symmetry is due to the conserved current $j^{\mu}=\frac{1}{2\pi}\epsilon^{\mu\nu}\partial_{\nu}\varphi$ in $2\mathrm{d}$. We denote this second $U(1)$ symmetry by $U_{wind}(1)$. Under the T-duality $(R\to 2/R)$ we have the correspondence between the symmetries: $U_{shift}(1)\to U_{wind}(1)$ and $U_{wind}(1)\to U_{shift}(1)$. In the present work we focus on $U_{wind}(1)$ at arbitrary compactification radius $R$.

The $n=2$ R\'enyi entropy for two disjoint intervals case was first studied in \cite{furukawa2009mutual}, and this result was generalised to arbitrary integer values of $n$ in \cite{calabrese2009entanglement2}. The R\'enyi entropies for multi-interval case was studied in \cite{coser2014renyi, headrick2013bose}. The $n^{th}$ moment of the reduced density matrix for arbitrary integer values of $n$ and $N$ is given by the generalised expression 
\begin{equation}\label{eq3.2}
Z_{N,n}=c_{N,n}\prod_{i=1}^{N}\left(\ell_i^{\frac{1}{6}\left(\frac{1}{n}-n\right)}\right)\prod_{j>i=1}^{N}\left(y_{ij}^{\frac{1}{6}\left(\frac{1}{n}-n\right)}\right)\mathcal{F}(x_1,\cdots,x_N),
\end{equation}
where $c_{N,n}$ is a non-universal constant and $y_{ij}$ are the cross-ratios
\begin{equation} \label{eq3.3}
y_{ij}=\frac{\left(u_j-v_i\right)\left(v_j-u_i\right)}{\left(u_j-u_i\right)\left(v_j-v_i\right)}.
\end{equation}
The factor $\mathcal{F}_{N,n}$ comes from the local conformal algebra (i.e. it is not fixed by the global conformal invariance) and is a function of the cross-ratios $x_i$ (introduced later in eq.\eqref{4.4}). It is given by
\begin{equation}\label{eq3.4}
\mathcal{F}(x_1,\cdots,x_N)=\frac{\Theta\left(0|T\right)}{\left|\Theta\left(0|\tau\right)\right|^2}.
\end{equation}
In the above equation, $\Theta$ is the Riemann Siegel theta function. A $k-$dimensional theta function is defined as
\begin{equation} \label{eq3.5}
\Theta\left[\begin{array}{l}
\boldsymbol{\varepsilon} \\
\boldsymbol{\delta}
\end{array}\right](\boldsymbol{u} \mid \Omega) \equiv \sum_{\boldsymbol{m} \in \mathbb{Z}^{k}} e^{i \pi(\boldsymbol{m}+\boldsymbol{\varepsilon})^t \cdot \Omega \cdot (\boldsymbol{m}+\boldsymbol{\varepsilon})+2 \pi i(\boldsymbol{m}+\boldsymbol{\varepsilon})^t \cdot(\boldsymbol{u}+\delta)},
\end{equation}
where the characteristics $\boldsymbol{\varepsilon}$, $\boldsymbol{\delta} \in \left(\mathbb{Z}/2\right)^k$ and $\boldsymbol{u} \in \mathbb{C}^k$. In eq.\eqref{eq3.5} $\Omega$ is a $k\times k$ symmetric matrix with a positive definite imaginary part. When $\varepsilon=\delta=0$, the theta function in eq.\eqref{eq3.5} is denoted  $\Theta(\boldsymbol{u}|\Omega)$.

In eq.\eqref{eq3.4}, $\tau$ is the Riemann period matrix associated with the Riemann surface of the $n^{th}$ moment of the reduced density matrix. It is a $g\times g$ matrix, where $g$ is the genus of the Riemann surface. Let us write $\tau$ as
\begin{equation} \label{eq3.6}
\tau=\mathcal{R}+i\mathcal{I},
\end{equation} 
where $\mathcal{R}$ and $\mathcal{I}$ are $g\times g$ real symmetric matrices. Then the matrix $T$ in eq.\eqref{eq3.4} is given by
\begin{equation} \label{eq3.7}
T(x)=\begin{pmatrix}
iK\mathcal{I} & \mathcal{R}\\
\mathcal{R} & i\mathcal{I}/K
\end{pmatrix},
\end{equation}
where $K$ is the Luttinger parameter introduced earlier. We note from eq.\eqref{eq3.2}, that $Z_{N,n}$ is invariant under $K\to1/K$, as we would expect from the T-duality of compact boson. It has been noted in \cite{headrick2013bose}, that $Z_{N,n}$ at self-dual radius matches the corresponding $Z_{N,n}$ of the Dirac fermion when $\mathcal{R}$ in eq.\eqref{eq3.6} vanishes. This happens in particular for all values of $n$ when $N=2$, and similarly for all values of $N$ when $n=2$. It should also be noted that the factor $\mathcal{F}_{N,n}$ becomes unity at self-dual radius when $\mathcal{R}$ vanishes.
\section{Multi-Charged moments} \label{section4}
In this section, we obtain the multi-charged and charged moments of free compact boson for an arbitrary number of disjoint intervals.

The multi-charged moments of free compact boson in two disjoint intervals case for R\'enyi entropy was first obtained in \cite{ares2022multi}, these moments were then used to study the symmetry resolved R\'enyi entropy and mutual information. In \cite{gaur2023charge}, these quantities were obtained for R\'enyi negativity to study the symmetry-resolved R\'enyi negativity in the same settings.

The flux generating vertex operators $\mathcal{V}_{\alpha}$ in \eqref{eq2.12} for free compact boson are just the boson vertex operators \cite{goldstein2018symmetry}
\begin{equation} \label{eq4.1}
\mathcal{V}_{\alpha}(z)=e^{i\alpha\frac{\varphi(z)}{2\pi}},
\end{equation}
with the scaling dimension $h_\alpha^{\mathcal{V}}$ 
\begin{equation} \label{eq4.2}
h_\alpha^{\mathcal{V}}=\left(\frac{\alpha}{2\pi}\right)^2\frac{K}{2}.
\end{equation}
To simplify our calculations we first use the global conformal invariance to map the points $u_{1}\to 0$, $u_{N}\to 1$, and $v_{N}\to \infty$ using the map
\begin{equation} \label{eq4.3}
w=\frac{\left(z-u_{1}\right)\left(v_{N}-u_{N}\right)}{\left(v_{N}-z\right)\left(u_N-u_{1}\right)}.
\end{equation}
Let's denote the image of all the boundary points under this map as
\begin{equation}\label{4.4}
x_{2j-2}=\frac{\left(u_{j}-u_{1}\right)\left(v_{N}-u_{N}\right)}{\left(v_{N}-u_{j}\right)\left(u_N-u_{1}\right)}, \qquad x_{2j-1}=\frac{\left(v_{j}-u_{1}\right)\left(v_{N}-u_{N}\right)}{\left(v_{N}-v_{j}\right)\left(u_N-u_{1}\right)},
\end{equation}
where $j\in \{1,2,\cdots N\}$. We have the order $0<x_1<x_2\cdots<x_{2N-3}<1$ and we also keep in mind the notations $x_0=0$, $x_{2N-2}=1$, and $x_{2N-1}=\infty$. We must now compute the correlation function for the Vertex operators at $x_i$ on the Riemann surface $\Sigma_{N,n}(\boldsymbol{x})$ (see Appendix \ref{A}), i.e. $\left\langle\prod_{i=1}^N\mathcal{V}_{\alpha_j}(x_{2j-2})\mathcal{V}_{-\alpha_j}(x_{2j-1})\right\rangle$. The Riemann surface $\Sigma_{N,n}(\boldsymbol{x})$ has the genus $g=(N-1)\times (n-1)$. The correlation functions of vertex operators have been studied extensively in the string theory literature \cite{verlinde1987chiral, eguchi1987chiral}. A general $M-$point correlation function is given by
\begin{equation} \label{4.5}
\begin{split}
&\left\langle \prod_{i=1}^{M}\mathcal{V}_{\beta_i}(z_i)\right\rangle_{\Sigma_{N,n}(\boldsymbol{x})}=\\
&\hspace{0.8in}\prod_{1\leq i<i'\leq M}\left| E(z_i,z_{i'})e^{-\pi \mathrm{Im}|\boldsymbol{w}(z_i)-\boldsymbol{w}(z_{i'})|^t\cdot\mathrm{Im}|\tau(x)^{-1}|\cdot\mathrm{Im}|\boldsymbol{w}(z_i)-\boldsymbol{w}(z_{i'})|}\right|^{\frac{\beta_i\beta_{i'}K}{2\pi^2}}.
\end{split}
\end{equation}
The map $\boldsymbol{w}$ is a $g$ dimensional vector and is known as the Abel-Jacobi map. It is defined from the Riemann surface $\Sigma_{N,n}(\boldsymbol{x})$ to its Jacobian torus $J\left(\Sigma_{N,n}(\boldsymbol{x})\right)$. For the Riemann surface $\Sigma_{N,n}(\boldsymbol{x})$, we define its Jacobian lattice $\Lambda=\mathbb{Z}^g+\tau(\boldsymbol{x})\mathbb{Z}^g$, where $\tau(\boldsymbol{x})$ (given by eq.\eqref{eqA.10}) is the period matrix of $\Sigma_{N,n}(\boldsymbol{x})$. The Jacobian torus is then defined as the quotient space $\mathbb{C}^g/\Lambda$. The Abel-Jacobi map $\boldsymbol{w}$ is given by
\begin{equation} \label{4.6}
w_{j,r}(z)=\int_0^z \mathrm{d}z'\nu_{j,r}(z')\; \mod{\Lambda},
\end{equation}
where $j\in(1,2,\cdots,(N-1))$, and $k\in(1,2,\cdots,(n-1))$ and we have chosen $z=0$ as the reference point. The quantities $\nu_{j,r}(z)$ are the normalised holomorphic differentials given by eq.\eqref{eqA.9} in Appendix \ref{A}.

The quantity $E(z_i,z_{i'})$ is the Prime form of the Riemann surface $\Sigma_n(\boldsymbol{x})$ and is given by \cite{fay2006theta, mumford2007tata}
\begin{equation} \label{eq4.7}
E(z_i,z_{i'})=\frac{\Theta_{\boldsymbol{\Delta}}\left(\boldsymbol{w}(z_i)-\boldsymbol{w}(z_{i'})|\tau(x)\right)}{h_{\boldsymbol{\Delta}}(z_i)h_{\boldsymbol{\Delta}}(z_{i'})},
\end{equation}
where ${\boldsymbol{\Delta}}=({\boldsymbol{\varepsilon}},{\boldsymbol{\delta}})$, where $\boldsymbol{\varepsilon}$, and $\boldsymbol{\delta}$ were defined below eq.\eqref{eq3.5}. However, here $\boldsymbol{\Delta}$ must be a non-singular odd half characteristic. The prime form $E(z_i,z_{i'})$ is independent of the choice of $\boldsymbol{\Delta}$. The quantity $h_{\boldsymbol{\Delta}}(z_i)$ is a holomorphic 1-form and is given by
\begin{equation} \label{eq4.8}
h_{\boldsymbol{\Delta}}(z_i)=\left(\sum_{j=1}^{N-1}\sum_{r=1}^{n-1}\nu_{j,r}(z_i)\partial_{u_{j,r}}\Theta_{\boldsymbol{\Delta}}(\boldsymbol{u}|\tau)|_{u=0}\right)^{\frac{1}{2}}.
\end{equation}
In our case, $z_i$ are the branch points and the normalised holomorphic differentials in eq.\eqref{eq4.8} are singular. This makes the vertex operator correlation function in eq.\eqref{4.5} ill-defined. This issue is resolved by introducing the regularised vertex operators $\mathcal{V}^{(*)}_{\alpha_i}(z_i)$ \cite{ares2022multi}. The normalised holomorphic differentials $\nu_{j,r}(z_i)$ exhibit the leading order singular behaviour as $z\to z_i$
\begin{equation} \label{eq4.9}
\nu_{j,r}(z_i)\sim \lim_{\epsilon\to 0}\epsilon^{-\frac{n-1}{n}}\nu^{(*)}_{j,r}(z_i),
\end{equation}
where $\nu^{(*)}_{j,r}(z_i)$ is non-singular at the branch points. Using $\nu^{(*)}_{j,r}(z_i)$ in eq.\eqref{eq4.8}, and eq.\eqref{eq4.7} we may define the regularised holomorphic $1-$forms and regularised prime forms
\begin{align} 
&h^{(*)}_{\boldsymbol{\Delta}}(z_i)=\left(\sum_{j=1}^{N-1}\sum_{r=1}^{n-1}\nu^{(*)}_{j,r}(z_i)\partial_{u_{j,r}}\Theta_{\boldsymbol{\Delta}}(\boldsymbol{u}|\tau)|_{u=0}\right)^{\frac{1}{2}}, \label{eq4.10}\\
&E^{(*)}(z_i,z_{i'})=\frac{\Theta_{\boldsymbol{\Delta}}\left(\boldsymbol{w}(z_i)-\boldsymbol{w}(z_{i'})|\tau(x)\right)}{h^{(*)}_{\boldsymbol{\Delta}}(z_i)h^{(*)}_{\boldsymbol{\Delta}}(z_{i'})}. \label{eq4.11}
\end{align}
This leads us to define the regularised Vertex operators $\mathcal{V}^{(*)}_{\alpha_i}(z_i)$
\begin{equation} \label{eq4.12}
\mathcal{V}^{(*)}_{\alpha_i}(z_i)=\lim_{\epsilon\to 0}\left(\kappa_{n}\epsilon^{\frac{n-1}{n}}\right)^{2h_{\alpha_i}}\mathcal{V}_{\alpha_i}(z+\epsilon),
\end{equation}
where $\kappa_n$ is a surface-dependent constant. The constant $\kappa_n$ will be fixed later, but we will find that it is independent of $N$. The correlation function of the regularised vertex operators at branch points is well defined, and is given by
\begin{equation} \label{eq4.13}
\begin{split}
&\left\langle \prod_{i=0}^{2N-1}\mathcal{V}^{(*)}_{\beta_i}(z_i)\right\rangle_{\Sigma_{N,n}(\boldsymbol{x})}=\\
&\hspace{0.8in}\prod_{0\leq i<i'\leq 2N-1}\left| E^{(*)}(z_i,z_{i'})e^{-\pi \mathrm{Im}|\boldsymbol{w}(z_i)-\boldsymbol{w}(z_{i'})|^t\cdot\mathrm{Im}|\tau(x)^{-1}\cdot|\mathrm{Im}|\boldsymbol{w}(z_i)-\boldsymbol{w}(z_{i'})|}\right|^{\frac{\beta_i\beta_{i'}K}{2\pi^2}},
\end{split}
\end{equation}
where $z_{i}=x_{i}$. The flux is to be identified as $\beta_{2j-2}=\alpha_{j}$, and $\beta_{2j-1}=-\alpha_{j}$ in the equation above, where $j\in\{1,2,\cdots,N\}$.

In the rest of this section, we will evaluate this correlation function for arbitrary values of $N$ and $n$. We first revisit the two disjoint intervals case, we will use a different canonical homology basis than \cite{ares2022multi}. We then extend these results to $N>2$.
\subsection{Two-interval case}
The normalised holomorphic differential $\boldsymbol{\nu}$ in the case of two disjoint intervals for arbitrary values of $n$ is evaluated by using eq.\eqref{eqA.7} in eq.\eqref{eqA.5}, and eq.\eqref{eqA.9}. It is given by
\begin{equation} \label{eq4.14}
\nu_r=-\frac{1}{n\pi}\sum_{s=1}^{n-1}\frac{e^{-i\pi(r-3)\frac{s}{n}}\sin{\frac{\pi rs}{n}}}{F_{s/n}(x)}\left(z(z-1)\right)^{-\frac{s}{n}}(z-x)^{-1+\frac{s}{n}},
\end{equation}
where $F_{s/n}(x)\equiv {}_{2}F_{1}\left(s/n,1-s/n;1;x\right)$ is the hypergeometric function. The Abel-Jacobi map of the branch points is computed by using eq.\eqref{eq4.14} in eq.\eqref{4.6},
\begin{align}
w_r(0) &=0, \label{eq4.15} \\
w_r(x) &=1-\frac{r}{n}, \label{eq4.16}\\
w_r(1) &=1-\frac{r}{n}+iu_{r},\label{eq4.17} \\
w_r(\infty) &=iu_r,\label{eq4.18}
\end{align}
where $u_r$ is given by
\begin{equation} \label{eq4.19}
u_r=\frac{1}{n}\sum_{s=1}^{n}\sin\left(\pi(r-2)\frac{s}{n}\right)\frac{\sin{\frac{\pi rs}{n}}}{\sin{\frac{\pi s}{n}}}\frac{F_{s/n}(1-x)}{F_{s/n}(x)}.
\end{equation}
As noted earlier the normalised holomorphic differentials are singular at the branch points, and the corresponding non-singular $\nu^{(*)}_r$ is given by
\begin{equation} \label{eq4.20}
\nu^{(*)}_r(z)=\left\{
\begin{array}{ll}
x^{-1/n}Q_{r,n}(x), & z=0,\\
\left(x(1-x)\right)^{-1/n}e^{\frac{-i2\pi(r-3)}{n}}Q_{r,n}(x), & z=x,\\
(1-x)^{-1/n}e^{i\frac{\pi}{n}}Q_{r,n}(x), & z=1,\\
-e^{\frac{-i2\pi(r-3)}{n}}Q_{r,n}(x), & z=\infty,
\end{array}
\right.
\end{equation}
where $Q_{r,n}=\frac{e^{\frac{i\pi(r-3)}{n}}\sin{\frac{\pi r}{n}}}{n\pi F_{1/n}(x)}$. Finally, the regularised prime forms are evaluated using eq.\eqref{eq4.20} in eq.\eqref{eq4.10} and eq.\eqref{eq4.11}. These prime forms were conjectured to be simple algebraic functions in \citep{ares2022multi},
\begin{align}
|E^{(*)}(x,0)|&=n x^{1/n}, \label{eq4.21}\\
|E^{(*)}(1,\infty)|&=n, \label{eq4.22} \\
|p(0,x,1,\infty)|&=(1-x)^{1/n}, \label{eq4.23}
\end{align} 
where $p(0,x,1,\infty)$ is a cross-ratio function on the Riemann surface and is given by
\begin{equation} \label{eq4.24}
p(x_{i},x_{j};x_{k},x_{l})=\frac{E(x_{j},x_{k})E(x_{i},x_{l})}{E(x_{j},x_{l})E(x_{i},x_{k})}=\frac{E^{(*)}(x_{j},x_{k})E^{(*)}(x_{i},x_{l})}{E^{(*)}(x_{j},x_{l})E^{(*)}(x_{i},x_{k})}.
\end{equation}
We numerically checked these conjectures for a few cases in Appendix \ref{B} and found good agreement. The multi-charged moments are found by using eq.\eqref{eq4.15}-\eqref{eq4.18}, and eq.\eqref{eq4.21}-\eqref{eq4.23} in eq.\eqref{eq4.13}, and inverting the conformal transformation in eq.\eqref{eq4.3}. In the limit of large separation between the two intervals, the multi-charged moments are just the product of the charged moments of two single intervals. This leads us to set the structure constant $\kappa_n=n$. The multi-charged moments are found to be
\begin{equation}
Z_{2,n}(\alpha_1,\alpha_2)= c_{2,n;\alpha_1,\alpha_2}{\ell}_1^{-\frac{\alpha_1^2 K}{2\pi^2n}}{\ell}_2^{-\frac{\alpha_2^2 K}{2\pi^2n}}{(1-x)}^{-\frac{\alpha_1\alpha_2 K}{2\pi^2n}}Z_{2,n},
\end{equation}
where $c_{2,n;\alpha_1,\alpha_2}$ is the non-universal constant and $Z_{2,n}$ is given by eq.\eqref{eq3.2} without the non-universal constant. As expected, we find agreement with the known results of two-disjoint intervals.
\subsection{Multi-interval case (N>2)}
Let us first show that the exponential term in eq.\eqref{eq4.13} evaluates to unity. We show this by noting that if the condition
\begin{equation} \label{eq4.25}
\mathrm{Img}\left(\int_{x_{2j-2}}^{x_{2j-1}}\mathrm{d}z\nu_{k,s}(z)\right)=0\; \mod{\Lambda}\qquad \forall j\in\{1,2,\cdots,N-1\},
\end{equation}
holds then the exponent in eq.\eqref{eq4.13} evaluates to zero. To argue this, consider when $i=2j-2$, and $i^{'}=2j-1$ in eq.\eqref{eq4.13}, the corresponding exponent vanishes if eq.\eqref{eq4.25} holds. Now, the exponent corresponding to $i=2j-2$, and $i^{'}=2k-1$, $j\neq k$, is just the negative of when $i=2j-2$, and $i^{'}=2k-1$. Finally, we note that $w_{j,r}(\infty)$ is given by
\begin{equation} \label{eq4.26}
w_{j,r}(\infty)=\sum_{k=1}^{N-1}\int_{x_{2j-1}}^{x_{2j}}\mathrm{d}z\nu_{j,r}(z).
\end{equation}  
Then from eq.\eqref{eq4.25}, we have $\mathrm{Img}\left(\boldsymbol{w}(\infty)-\boldsymbol{w}(1)\right)=0$, which completes our argument.
Now to demonstrate that eq.\eqref{eq4.25} holds here, first we write from eq.\eqref{eqA.5}, and eq.\eqref{eqA.9}
\begin{equation} \label{eq4.27}
\int_{x_{2j-2}}^{x_{2j-1}}\mathrm{d}z\nu_{k,r}(z)=\sum_{l,s}\left(\mathcal{A}^{-1}\right)^{k,l}_{r,s}\frac{\left(\mathcal{A}_{s,1}^{l,j}-\mathcal{A}_{s,1}^{l,j-1}\right)}{1-e^{-i\frac{2\pi s}{n}}}.
\end{equation}
The matrix $\mathcal{A}$ may be decomposed as follows \cite{coser2014renyi}
\begin{equation} \label{eq4.28}
\mathcal{A}_{r,s}^{j,k}=\mathcal{M}_{r,s}\left(\mathcal{A}_r\right)_{j,k},
\end{equation}
where the $(n-1)\times(n-1)$ matrix $\mathcal{M}$ and the $(N-1)\times(N-1)$ matrices $\mathcal{A}_r$ are given by
\begin{align} 
\mathcal{M}_{r,s}&=e^{i\frac{2\pi rs}{n}} \label{eq4.29}\\
\left(\mathcal{A}_r\right)_{j,k}&=e^{-i\frac{2\pi r}{n}}\left(e^{-i\frac{2\pi r}{n}}-1\right)\sum_{l=1}^k\int_{x_{2l-2}}^{x_{2l-1}}\mathrm{d}z w_{j,r}(z). \label{eq4.30}
\end{align} 
Note that our definition of $\mathcal{M}$ and $\mathcal{A}_r$ differs slightly from the reference. It then follows that $\mathcal{A}^{-1}$ also admits a similar decomposition, $\left(\mathcal{A}^{-1}\right)_{r,s}^{j,k}=\mathcal{M}^{-1}_{r,s}\left(\mathcal{A}^{-1}_s\right)_{j,k}$, were we have the relations $\mathcal{M}^{-1}\cdot\mathcal{M}=\mathbb{I}$ and $\mathcal{A}^{-1}_r\cdot\mathcal{A}_r=\mathbb{I}$. The matrix $\mathcal{M}^{-1}$ is given by $\left(\mathcal{M}^{-1}\right)_{r,s}=\frac{1}{n}\left(e^{-i\frac{2\pi rs}{n}}-1\right)$. Using these decompositions of $\mathcal{A}$, and $\mathcal{A}^{-1}$ in eq.\eqref{eq4.27} we obtain
\begin{equation}\label{eq4.31}
\int_{x_{2j-2}}^{x_{2j-1}}\mathrm{d}z\nu_{k,r}(z)=\left(1-\frac{r}{n}\right)\left(\delta_{k,j}-\delta_{k,j-1}\right).
\end{equation}
This shows that the condition in eq.\eqref{eq4.24} is satisfied.

We now proceed to study the normalised holomorphic differentials at the branch points. As discussed earlier they are singular at the branch points. The quantities $\nu^{(*)}_{j,k}$ at the branch points are given by
\begin{align}
\nu^{(*)}_{k,r}(x_{2j-2})&=P_{r,n}\left(\sum_{l=1}^{n-1}\left(\mathcal{A}_{n-1}^{-1}\right)_{k,l}x_{2j-2}^{l-1}\right)\prod_{l=1,\neq j}^{N}\left(x_{2j-2}-x_{2l-2}\right)^{-1+\frac{1}{n}}\prod_{l=1}^{N-1}\left(x_{2j-2}-x_{2l-1}\right)^{-\frac{1}{n}} \label{eq4.32}\\
\nu^{(*)}_{k,r}(x_{2j-1})&=\frac{P_{r,n}}{e^{i\frac{2\pi(r-3)}{n}}}\left(\sum_{l=1}^{n-1}\left(\mathcal{A}_{1}^{-1}\right)_{k,l}x_{2j-1}^{l-1}\right)\prod_{l=1}^{N}\left(x_{2j-1}-x_{2l-2}\right)^{-\frac{1}{n}}\prod_{l=1,\neq j}^{N-1}\left(x_{2j-1}-x_{2l-1}\right)^{-1+\frac{1}{n}} \label{eq4.33}\\
\nu^{(*)}_{k,r}(\infty)&=\frac{P_{r,n}}{e^{i\frac{2\pi(r-3)}{n}}}(\left(\mathcal{A}_{1}^{-1}\right)_{k,n-1}, \label{eq4.34}
\end{align}
where $P_{r,n}=\left(e^{i\frac{\pi(r-3)}{n}}\sin{\frac{\pi r}{n}}\right)/\left(\sin{\frac{\pi}{n}}\right)$ and $j\neq N$ in eq.\eqref{eq4.33}. The matrices $\mathcal{A}_1^{-1}$, and $\mathcal{A}_{n-1}^{-1}$ are the inverses of the matrices $\mathcal{A}_1$, and $\mathcal{A}_{n-1}$ given by eq.\eqref{eq4.30}. The regularised prime-forms are again given by using $\nu^{(*)}_{k,r}$ in eq.\eqref{eq4.10}, and eq.\eqref{eq4.11}. Following the two disjoint intervals case, we conjecture that the relatively complicated regularised prime forms are given by the simple algebraic functions
\begin{align}
\left| E^{(*)}\left(x_{2j-2},x_{2j-1}\right)\right|&=n\left(x_{2j-1}-x_{2j-2}\right)^{\frac{1}{n}},\label{eq4.35}\\
\left| p\left(x_{2j-2},x_{2j-1};x_{2k-2},x_{2k-1}\right)\right|&=\left(\frac{(x_{2k-1}-x_{2j-2})(x_{2k-2}-x_{2j-1})}{(x_{2k-1}-x_{2j-1})(x_{2k-2}-x_{2j-2})}\right)^{\frac{1}{n}},\;(k>j),\label{eq4.36}\\
\left| E^{(*)}\left(1,\infty\right)\right|&=n,\label{eq4.37}\\
\left| p\left(x_{2j-2},x_{2j-1};1,\infty\right)\right| &=\left(\frac{(1-x_{2j-2})}{(1-x_{2j-1})}\right)^{\frac{1}{n}}.\label{eq4.38}
\end{align} 
where $k,j\in(1,2,\cdots,N-1)$. The cross ratio functions $p$ are defined in eq.\eqref{eq4.24}.

\begin{figure}
\centering 
\includegraphics[width=1\textwidth]{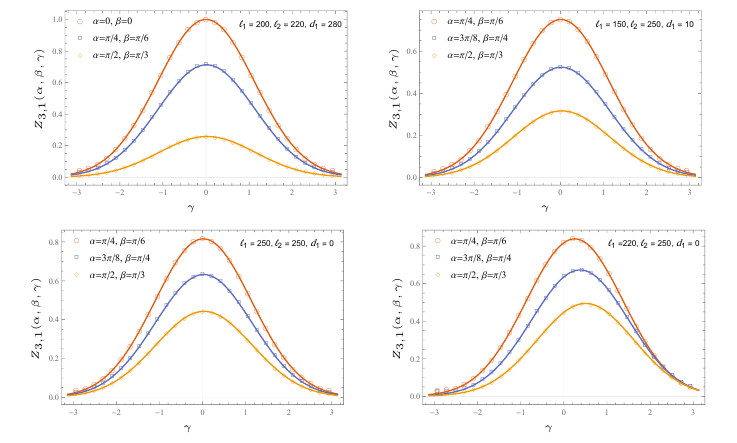}
\caption{\label{fig:iii} Plots for the multi-charged moments. The continuous lines are plots of $Z_{3,1}(\alpha,\beta,\gamma)$ (given by using $N=3$ and $n=1$ in eq.\eqref{eq4.40} for the top two. For the bottom plot $d_1\to 0$ limit is taken in eq.\eqref{eq4.40}) for $K=1$ as a function of $\gamma$ at different values of $\alpha$ and $\beta$. The discrete points are the plots of the numerically evaluated multi-charged moments for the tight-binding model given in Appendix \ref{D}.} 
\end{figure}
\begin{figure}
\centering 
\includegraphics[width=1\textwidth]{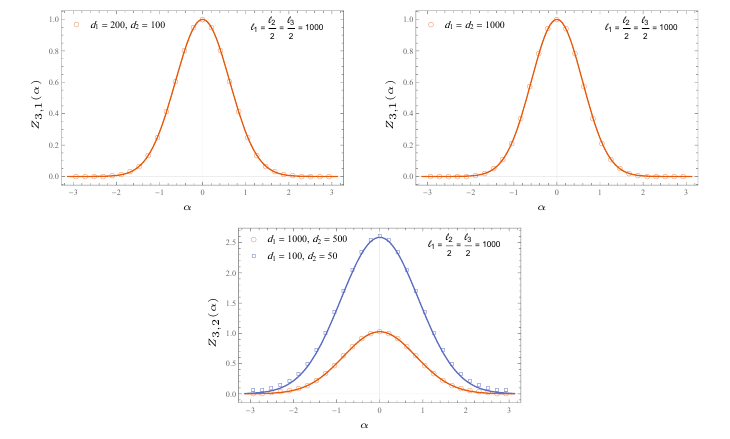}
\caption{\label{fig:iv} Plots for the charged moments. The continuous lines are plots of $Z_{3,1}(\alpha)$ and $Z_{3,2}(\alpha)$ (given by using $N=3$, and $n=1,2$ in eq.\eqref{eq4.41} for the top two and bottom plots) for $K=1$ as a function of $\alpha$. The discrete points are the plots of the numerically evaluated charged moments for the tight-binding model given in Appendix \ref{D}.} 
\end{figure}
We checked these conjectures numerically for a few cases in Appendix \ref{B} and found good agreement. There, however, arises a subtle issue of Riemann zeros in the numerical implementation of prime forms, this issue has been also discussed in Appendix \ref{B}. The regularised vertex operator correlation function given by eq.\eqref{eq4.13} is evaluated to be
\begin{equation} \label{eq4.39}
\begin{split}
\left\langle \prod_{i=0}^{2N-1}\mathcal{V}^{(*)}_{\beta_i}(z_i)\right\rangle_{\Sigma_{N,n}(\boldsymbol{x})}=\left(\frac{\kappa}{n}\right)^{2\sum_i h_{\alpha_i}}&\prod_{j=1}^{N-1}(x_{2j-1}-x_{2j-2})^{-\frac{K\alpha_j^2}{2\pi^2 n}}\left(\frac{1-x_{2j-1}}{1-x_{2j-2}}\right)^{-\frac{K\alpha_j\alpha_N}{2\pi^2 n}}\\
&\prod_{j<k=1}^{N-1}\left(\frac{(x_{2k-1}-x_{2j-2})(x_{2k-2}-x_{2j-1})}{(x_{2k-1}-x_{2j-1})(x_{2k-2}-x_{2j-2})}\right)^{-\frac{K\alpha_j\alpha_k}{2\pi^2n}}\end{split},
\end{equation}
where $z_i$, and $\beta_i$ on the left in the above equation are given below eq.\eqref{eq4.13}. Finally, after taking the global conformal transformation, we obtain the multi-charged moments
\begin{equation} \label{eq4.40}
Z_{N,n}(\alpha_1,\cdots,\alpha_N)=c_{N,n;\alpha_1,\cdots,\alpha_N}\left(\frac{\kappa_n}{n}\right)^{2\sum_i h_{\alpha_i}}\prod_{i=1}^{N}\ell_i^{-\frac{K\alpha_j^2}{2\pi^2 n}}\prod_{i<j=1}^{N}y_{ij}^{-\frac{K\alpha_i\alpha_j}{2\pi^2n}}Z_{N,n},
\end{equation}
where the cross-ratios $y_{ij}$ are given by eq.\eqref{eq3.3} and we have introduced the non-universal constant $c_{N,n;\alpha_1,\cdots,\alpha_N}$. As before, $Z_{N,n}$ is given by eq.\eqref{eq3.2} without the non-universal constant. 

In ref.\cite{coser2014renyi} it has been argued that in the limit of large separation between the intervals, the partition function $Z_{N,n}$ factorises into the product of the partition function of the individual interval $Z^{A_i}_{n}$, i.e. $Z_{N,n}$=$\prod_{i=1}^N Z^{A_i}_{n}$. We would expect this to hold for the multi-charged moments as well. This implies that we set $\kappa_n=n$. This also implies that the non-universal constant $c_{N,n;\alpha_i,\cdots,\alpha_N}$ also factorises into the non-universal constant of each interval for arbitrary interval lengths and distances.
The charged moments are similarly found to be
\begin{equation} \label{eq4.41}
Z_{N,n}(\alpha)=c_{N,n;\alpha}\prod_{i=1}^{N}\ell_i^{-\frac{K\alpha^2}{2\pi^2 n}}\prod_{i<j=1}^{N}y_{ij}^{-\frac{K\alpha^2}{2\pi^2n}}Z_{N,n},
\end{equation}
where we already used $\kappa_n=n$.

We may also study the limit of two intervals approaching each other in eq.\eqref{eq4.40} and eq.\eqref{eq4.41}. To do so, let's take the limit $d_i\to 0$ (i.e $v_i\to u_{i+1}$), in this limit we have
\begin{equation}
y_{i,i+1}\sim \lim_{\epsilon\to 0}\epsilon\frac{\ell_i+\ell_{i+1}}{\ell_i\ell_{i+1}},
\end{equation}
where the $\epsilon$ must now be absorbed into the UV cut-off. This leads to a different non-universal constant in eq.\eqref{eq4.40} and eq.\eqref{eq4.41}.

In Appendix \ref{C}, we also computed the multi-charged moments for the $2$d massless Dirac fermion in the same setting as well. We note that the multi-charged moments of compact boson at self-dual radius matches with that of massless Dirac fermions for the cases where the reduced density matrices of the two theory are known to match  \cite{headrick2013bose}. We also numerically checked our results for some of these cases against the tight-binding model, plots are shown in Figure \ref{fig:iii} and Figure \ref{fig:iv}. The non-universal constant for the tight-binding model has been found in \cite{bonsignori2019symmetry}. We see from the figures that we have a good numerical match. 
\section{Symmetry Resolved R\'enyi Entropy} \label{section5}
In this section, we obtain the Symmetry resolved R\'enyi entropy and $\mathcal{Z}(q_1,\cdots,q_N)$ by taking the Fourier transform of the multi-charged and charged moments.

To evaluate the Fourier transform of the multi-charged moments we need the functional form of the non-universal constant $c_{N,n;\alpha_1,\cdots,\alpha_N}$ in $\alpha_i$. In the last section, we argued that the non-universal constant for $N$ interval factorises into the product of the non-universal constant for the single interval. This allows us approximate the non-universal constant to the leading order in $\alpha_i$ as $c_{N,n;\alpha_1,\cdots,\alpha_N}\sim c_{N,n}\lambda_{n}^{-\frac{K}{2\pi^2n}\sum_{i=1}^N\alpha_i^2}$ \cite{xavier2018equipartition}. We will use this approximation in the rest of this section.

\begin{figure}
\centering 
\includegraphics[width=1\textwidth]{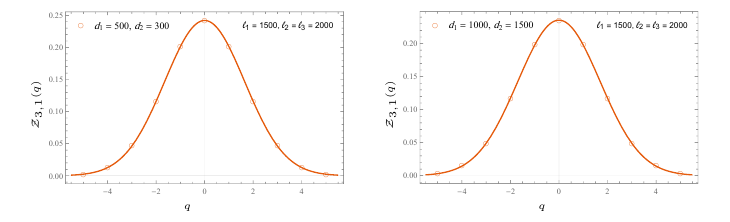}
\caption{\label{fig:v} Plots for the $\mathcal{Z}_{3,1}(q)$. The continuous lines are plots of $\mathcal{Z}_{3,1}(q)$ (given by using $N=3$ and $n=1$ in eq.\eqref{eq5.3}) for $K=1$ as a function of $q$. The discrete points are the plots of the numerically evaluated $\mathcal{Z}_{3,1}(q)$ for the tight-binding model given in Appendix \ref{D}.}
\end{figure}

\begin{figure}
\centering 
\includegraphics[width=1\textwidth]{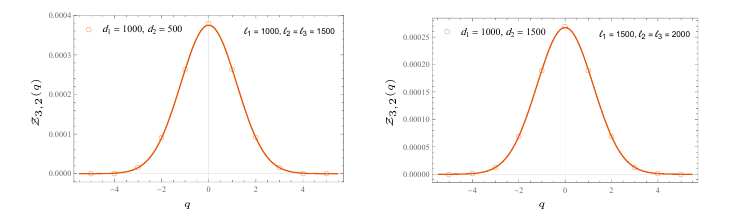}
\caption{\label{fig:vi} Plots for the $\mathcal{Z}_{3,2}(q)$. The continuous lines are plots of $\mathcal{Z}_{3,1}(q)$ (given by using $N=3$ and $n=2$ in eq.\eqref{eq5.3}) for $K=1$ as a function of $q$. The discrete points are the plots of the numerically evaluated $\mathcal{Z}_{3,2}(q)$ for the tight-binding model given in Appendix \ref{D}.}
\end{figure}

To evaluate $\mathcal{Z}_{N,n}(q_1,\cdots,q_N)$, let's first introduce $N\times N$ matrices $Y_n$ for brevity later, defined as
\begin{equation} \label{eq5.1}
\left(Y_n\right)_{i,j}=\left\{
\begin{array}{ll}
\log\left(\lambda_n{\ell}_i\right), & \text{when}\; j=i,\\
\frac{1}{2}\log\left(y_{ij}\right), & \text{when}\; j\neq i.
\end{array}
\right.
\end{equation}
We may write eq.\eqref{2.5} using eq.\eqref{eq4.41} and eq.\eqref{eq5.1}, using our approximation of the non-universal 
\begin{equation*}
\mathcal{Z}_{N,n}(\boldsymbol{q})=Z_{N,n}\prod_{i=1}^N\left(\frac{1}{2\pi}\int_{-\pi}^{\pi}\mathrm{d}\alpha_{i}\right)e^{-\frac{K}{2\pi^2 n}\boldsymbol{\alpha}^t\cdot Y_n\cdot\boldsymbol{\alpha}-i2\pi\boldsymbol{\alpha}^t\cdot\boldsymbol{q}},
\end{equation*}
where we have introduced $\boldsymbol{\alpha}^t=(\alpha_1,\cdots,\alpha_N)$ and $\boldsymbol{q}^t=(q_1,\cdots,q_N)$. To evaluate this integral, we first take the Gaussian approximation of the integral. Then from the standard techniques for evaluating multi-variable Gaussian integral, we obtain
\begin{equation} \label{eq5.2}
\mathcal{Z}_{N,n}(\boldsymbol{q})=Z_{N,n}\left(\frac{\pi n}{2K\det(Y_n)}\right)^\frac{N}{2} e^{-\frac{\pi^2 n}{2K}\boldsymbol{q}^t\cdot Y_n^{-1}\cdot\boldsymbol{q}}.
\end{equation} 
We note that the Luttinger parameter $K$ appears as an overall factor in the denominator of the exponent, promoting wider charge distribution. Similarly, we obtain $Z_{N,n}(q)$ in eq.\eqref{2.6} to be
\begin{equation} \label{eq5.3}
\mathcal{Z}_{N,n}(q)= Z_{N,n}\left(\frac{\pi n}{2K\Lambda_{N,n}}\right)^\frac{1}{2} e^{-\frac{\pi^2 n}{2K\Lambda_{N,n}}q^2},
\end{equation}
where for brevity we introduced $\Lambda_{N,n}=\log\left(\prod_{i=1}^{N}\lambda_n\ell_i\prod_{i<j=1}^{N}y_{ij}\right)$. The Luttinger parameter again appears as an overall factor in the denominator of the exponent and we see that the standard deviation is proportional to the Luttinger parameter. We have checked these results numerically against the tight-binding model in Figure \ref{fig:v} and Figure \ref{fig:vi}. Finally the symmetry resolved R\'enyi entropy is found to be
\begin{equation}\label{eq5.4}
\begin{split}
S_{N,n,q}=S_{N,n}+\frac{1}{2(1-n)}\left[n\log\left(\frac{2K\Lambda_{N,1}}{\pi^2}\right)-\log\left(\frac{2K\Lambda_{N,n}}{\pi^2}\right)\right]&+\frac{1}{2(1-n)}\log(n)\\-q^2\frac{\pi^2n}{2K(1-n)}&\left[\frac{N\left(\log\lambda_1-\log\lambda_n\right)}{\Lambda_{N,n}\Lambda_{N,1}}\right]
\end{split}
\end{equation} 
where $S_{N,n}$ is R\'enyi entropy of the free compact boson.

We see that to the leading order in $\ell_i$, we have the familiar result of the equipartition of symmetry resolved R\'enyi entropy \cite{goldstein2018symmetry, xavier2018equipartition}. This equipartition is broken by the terms of $O\left(1/\log^2(\ell_i)\right)$, similar results have been obtained for free fermions on a lattice in \cite{bonsignori2019symmetry}. Finally, the CFT result is given by
\begin{equation} \label{eq5.5}
S_{N,n,q}=S_{N,n}-\frac{1}{2}\log\left(\frac{2K}{\pi^2}\log\left(\prod_{i=1}^{N}{\ell}_i\prod_{i<j=1}^{N}y_{ij}\right)\right)+\frac{1}{2(1-n)}\log(n),
\end{equation}
We also note that the Luttinger parameter $K$ appears in the $O(1)$ terms. This generalises the symmetry-resolved R\'enyi entropy for the free compact boson for arbitrary disjoint intervals.
\section{Conclusion} \label{section6}
In this work, we evaluated the multi-charged moments and symmetry resolved R\'enyi entropy of free compact boson at arbitrary compactification radius for multiple disjoint intervals case. The symmetry resolved R\'enyi entropies were shown to have the familiar equipartition into the local charge sectors upto the leading order terms.

Free compact boson is the continuum theory of Luttinger liquids. The compactification radius $R$ of the free compact boson is related to the Luttinger parameter $K$ via $K=\sqrt{\frac{2}{R}}$. The charged and multi-charged moments of the theory are obtained by evaluating the correlation function of the flux-generating boson vertex operators placed at the branch points of the Riemann surface $\Sigma_{N,n}(\boldsymbol{x})$. The Riemann surface $\Sigma_{N,n}(\boldsymbol{x})$ is the associated replica space and has a genus $g=(N-1)\times(n-1)$. The multi-charged moments are given in terms of relatively complicated prime forms of the Riemann surface $\Sigma_{N,n}(\boldsymbol{x})$. These expressions are however simplified to algebraic functions of interval lengths and distances using the conjectures made in eq.\eqref{eq4.35}-\eqref{eq4.38}. Similar conjectures were first made in \citep{ares2022multi} for the two disjoint intervals case. The symmetry resolved R\'enyi entropy was then obtained by evaluating the Fourier transform of the charged moments. In Appendix \ref{C} the multi-charged moments of the massless Dirac field for multiple disjoint intervals were also evaluated. We found that the multi-charged moments of the self-dual compact boson and massless Dirac fermions match for the cases when the reduced density matrix of the two theories is known to match. Finally we also numerically checked out results for such cases against the tight-binding model. We found a good match between the analytical results and numerical evaluations.

Let's now discuss some future outlooks for the present work. R\'enyi entropies studied in this work only account for the entanglement between the subsystem $A$ and its complement $B$. To study the entanglement among the disjoint intervals mutual information measures are studied. These however are not measures of correlation, but still are interesting to study. The multi-charged moments obtained here are essential in the evaluation of symmetry-resolved mutual information \cite{parez2021exact}. Symmetry-resolved entanglement has been studied for Dirac fermions on the torus \cite{foligno2023entanglement}, and we believe that the results obtained here will prove to be useful in similar studies for free compact boson. Finally, the analytic results obtained here should be checked against the lattice models like $1$d spin chains.
\appendix
\section{Riemann Surfaces} \label{A}
In this section we review some topics in the theory of Riemann surfaces, particularly we focus on the normalised holomorphic differentials and the Riemann period matrix. For a detailed review, we refer the reader to \cite{fay2006theta, mumford2007tata} and to \cite{alvarez1987new, alvarez1986riemann} in the context of string theory.

We consider the singular Riemann surface $\Sigma_{N,n}(\boldsymbol{x})$, whose branch points are given by eq.\eqref{4.4}. The Riemann surface $\Sigma_{N,n}(\boldsymbol{x})$ has genus $g=(N-1)\times(n-1)$. This surface is parameterised by the curve
\begin{equation} \label{eqA.1}
y^n=\prod_{j=1}^N(z-x_{2j-2})\prod_{k=1}^{N-1}(z-x_{2j-1})^{n-1}.
\end{equation} 
The period matrix can be given in a canonical homology basis on the Riemann surface. To proceed with our aim, we must choose a canonical homology basis on the Riemann surface. Following \cite{enolski2004singular}, we first choose a set of auxiliary homology basis $\tilde{a}_{j,r}$ and, $\tilde{b}_{j,r}$, where $j\in\{1,2,\cdots,N-1\}$ and $r\in\{1,2,\cdots,n-1\}$. The non-contractible loop $\tilde{a}_{j,r}$ encloses the $j^{th}$ branch-cut on the $r^{th}$ sheet. The $\tilde{b}_{j,r}$ non-contractible loop goes through the $j^{th}$ branch-cut from the $r^{th}$ sheet to the $(r+1)^{th}$ sheet and returns to $r^{th}$ sheet through the $(j+1)^{th}$ branch-cut to close the loop. These loops do not form the canonical homology basis, since they do not satisfy the intersection conditions of the canonical homology basis
\begin{align}
{a}_{j,r}\circ {a}_{k,s}={b}_{j,r}\circ {b}_{k,s}&=0 \label{eqA.2}\\
{a}_{j,r}\circ {b}_{k,s}=-{b}_{j,r}\circ {a}_{k,s}&=\delta_{j,k}\delta_{r,s}.\label{eqA.3}
\end{align}
Canonical homology basis may be constructed from the auxiliary basis using the following relations
\begin{equation} \label{eqA.4}
{a}_{j,r}=\sum_{s=1}^{r}\tilde{a}_{j,s},\qquad 
{b}_{j,r}=\sum_{k=j}^{N-1}\tilde{b}_{k,r}.
\end{equation}
\begin{figure}
\centering 
\includegraphics[width=0.8\textwidth]{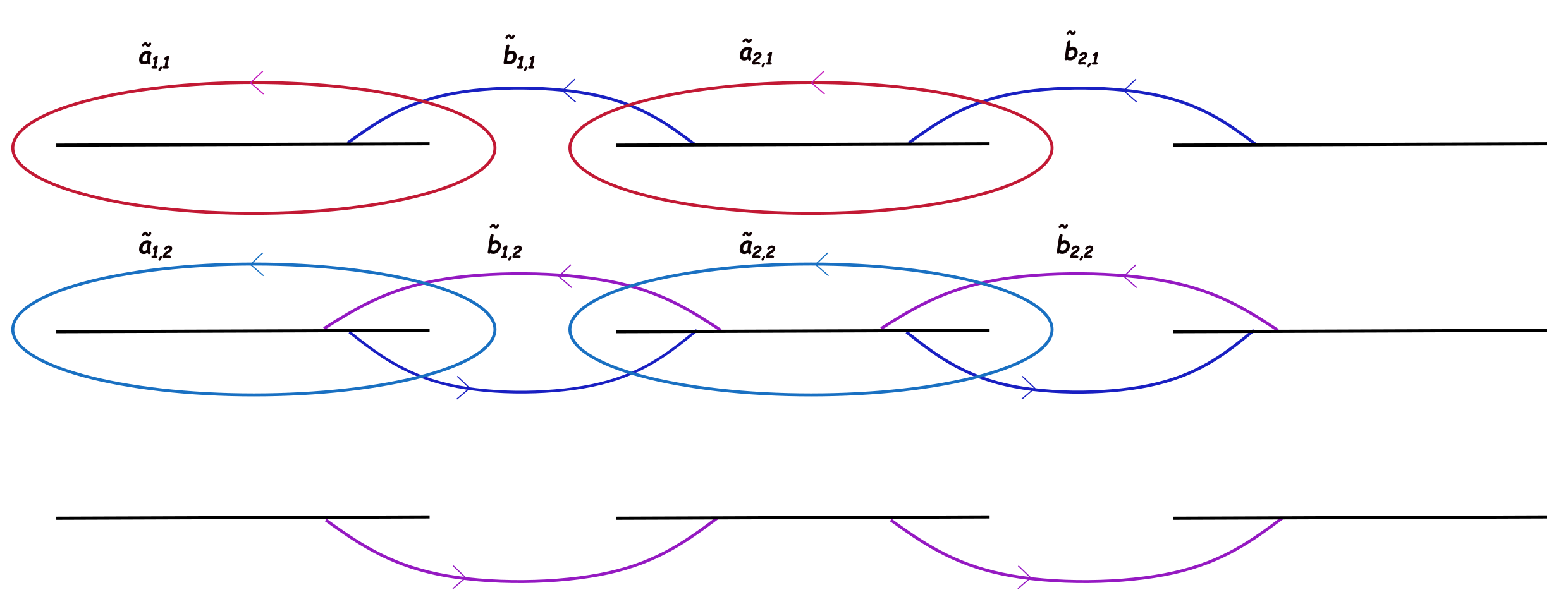}
\caption{\label{figA1} Auxiliary homology basis for the Riemann surface $\Sigma_{3,3}\left(\boldsymbol{x}\right)$}
\end{figure}
\begin{figure}
\centering 
\includegraphics[width=0.8\textwidth]{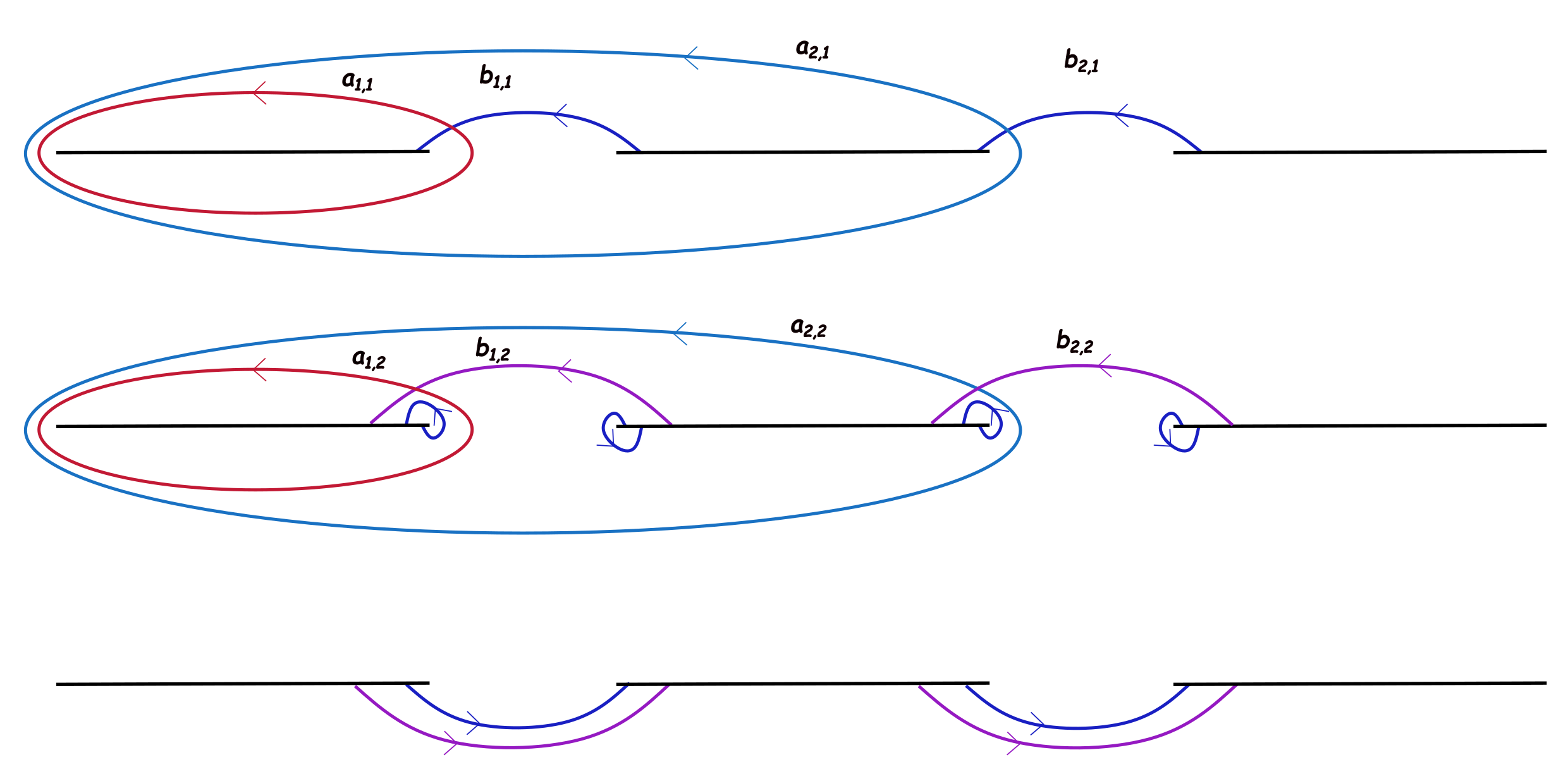}
\caption{\label{figA2} Canonical homology basis for the Riemann surface $\Sigma_{3,3}\left(\boldsymbol{x}\right)$}
\end{figure}

Having chosen our canonical homology basis, we now proceed to introduce the matrices $\mathcal{A}_{r,s}^{j,k}$ and $\mathcal{B}_{r,s}^{j,k}$. These matrices are defined as
\begin{align}
\mathcal{A}_{r,s}^{j,k}&=\oint_{a_{k,s}}\mathrm{d}\omega_{j,r},\label{eqA.5}\\
\mathcal{B}_{r,s}^{j,k}&=\oint_{b_{k,s}}\mathrm{d}\omega_{j,r}, \label{eqA.6}
\end{align}
where $\omega_{j,r}(z)$ are the basis of holomorphic differentials on $\Sigma_{N,n}(\boldsymbol{x})$. We define these holomorphic differentials as
\begin{equation} \label{eqA.7}
\omega_{j,r}(z)=z^{j-1}\prod_{k=1}^N(z-x_{2k-2})^{-\frac{r}{n}}\prod_{l=1}^{N-1}(z-x_{2l-1})^{1-\frac{r}{n}}.
\end{equation}
The normalised holomorphic differentials $\nu_{j,r}$ are defined by the normalisation condition
\begin{equation} \label{eqA.8}
\oint_{a_{j,r}}\nu_{k,s}=\delta_{j,k}\delta_{r,s}.
\end{equation} 
We may construct the normalised holomorphic differentials $\nu_{j,r}$ using the holomorhic differentials $\omega_{j,r}$, the normalised holomorphic differentials is given by
\begin{equation} \label{eqA.9}
\nu_{j,r}=\sum_{k,s}\left(\mathcal{A}^{-1}\right)_{r,s}^{j,k}\omega_{k,s}.
\end{equation}
Finally, the period matrix $\tau$ associated with the Riemann surface $\Sigma_{N,n}(\boldsymbol{x})$ is given by
\begin{equation} \label{eqA.10}
\tau_{r,s}^{j,k}=\oint_{b_{j,r}}\mathrm{d}\nu_{k,s}.
\end{equation}
The period matrix $\tau$ is the one which appears in the eq.\eqref{eq3.4}, and eq.\eqref{eq3.6}. 
\section{Prime forms and cross-rations on Riemann surface} \label{B}
In this appendix, we numerically check the conjectures made in eq.\eqref{eq4.21}-\eqref{eq4.23}, and eq.\eqref{eq4.35}-\eqref{eq4.38} for the regularised prime forms and cross-ratio function. We verified these conjectures numerically for a few cases and present the plots here.
\subsection*{N=2 case}
\begin{figure}[t]
\centering 
\includegraphics[width=0.9\textwidth]{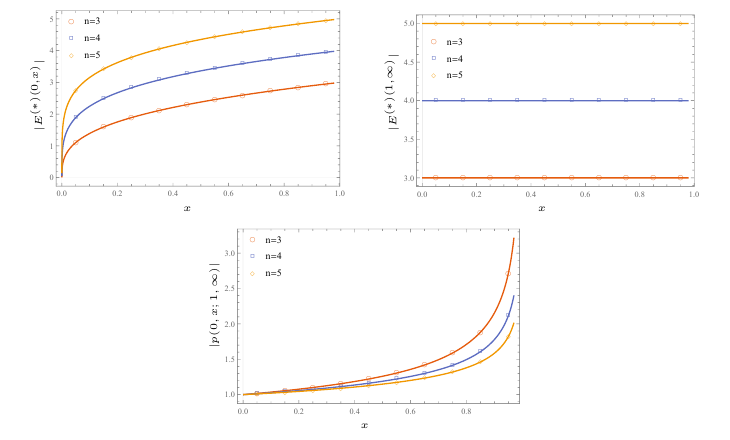}
\caption{\label{figB1} (N=2)Plots for the conjectures in eq.\eqref{eq4.21}-\eqref{eq4.23}. The solid lines are conjectured functions and the plotted points are numerically computed values of regularised prime forms and cross-ratio functions.}
\end{figure}

The conjectures for the $N=2$ case were first made in \citep{ares2022multi}, these conjectures were also proved for $n=2$ using the properties of Jacobi theta functions. Since we are using a different homology basis we numerically checked the conjectures for $n=3,4,5$. We used the odd-half characteristics  $\boldsymbol{\delta}=\boldsymbol{\varepsilon}=(1/2,0,\cdots,0)$. These checks are plotted in Figure \ref{figB1} and we see that we have a good match with the numerical results.
\subsection*{N>2 case}
In these cases, we encounter Riemann zeros at some branch points of the Riemann surface for $n=2$. Although the prime forms are independent of Riemann zeros, the numerical implementation of regularised prime forms becomes difficult. This problem is easily avoided by choosing the characteristics $\boldsymbol{\delta}$ and $\boldsymbol{\varepsilon}$ wisely for each regularised prime form and cross-ratio function. 

To give details, let's first discuss the Riemann theorem. It states that given a Abel-Jacobi map $\boldsymbol{w}_{P_0}(P)=\int_{P_0}^P\mathrm{d}\boldsymbol{\nu}$ defined from $\Sigma_{N,n}(\boldsymbol{x})\to J(\Sigma_{N,n}(\boldsymbol{x}))$, the Riemann theta functions either vanishes for all $P$ or has $g=(N-1)\times(n-1)$ zeros $\{P_i\}$. Mathematically, we write
\begin{equation} \label{eqB.1}
\Theta\left[\begin{array}{l}
\boldsymbol{\varepsilon} \\
\boldsymbol{\delta}
\end{array}\right](\boldsymbol{w}_{P_0}(P_i) \mid \tau)=0. 
\end{equation}
\begin{figure}[t]
\centering 
\includegraphics[width=0.9\textwidth]{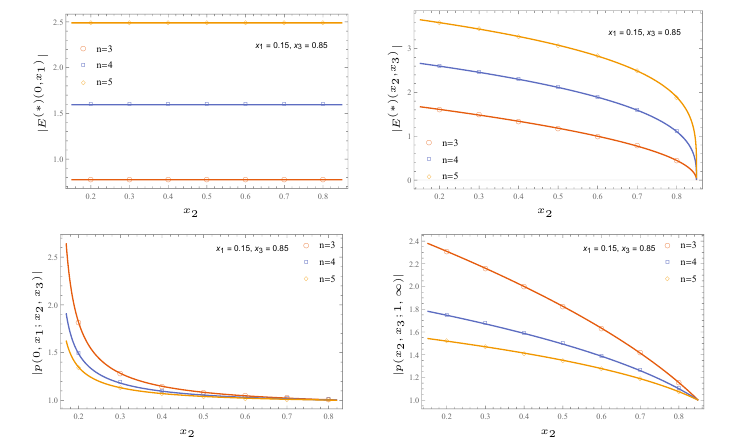}
\caption{\label{figB2} Plots for the conjectures in eq.\eqref{eq4.35}-\eqref{eq4.38} for $N=3$ and $n=2,3,4$. On the vertical axis, the solid lines are conjectured functions and the plotted points are numerically computed values of modified prime forms and cross-ratio functions.}
\end{figure}
\begin{figure}[t]
\centering 
\includegraphics[width=0.9\textwidth]{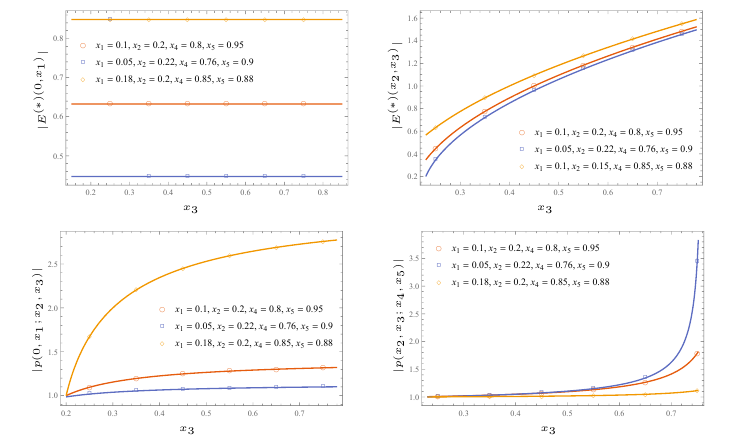}
\caption{\label{figB3} Plots for the conjectures in eq.\eqref{eq4.35}-\eqref{eq4.38} for $N=4$ and $n=2$. On the vertical axis, the solid lines are conjectured functions and the plotted points are numerically computed values of modified prime forms and cross-ratio functions.}
\end{figure}

Furthermore, the zeros satisfy $\sum_{i=1}^{g}\boldsymbol{w}_{P_0}(P_i)=\Delta$, where $\Delta$ is a constant vector depending only upon the homology basis and the reference point $P_0$. For odd-half characteristics, i.e $4\boldsymbol{\delta\cdot \varepsilon}=1\;\mathrm{mod}\;2$, $P_{0}$ is always a Riemann zero. Let $P$ be a Riemann zero, however, if we now choose different odd-half characteristics $\boldsymbol{\delta}'=\boldsymbol{\delta}+\boldsymbol{\tilde{\delta}}$ and $\boldsymbol{\varepsilon}'=\boldsymbol{\varepsilon}+\boldsymbol{\tilde{\varepsilon}}$, then form eq.\eqref{eq3.5}
\begin{equation} \label{eqB.2}
\Theta\left[\begin{array}{l}
\boldsymbol{\varepsilon'} \\
\boldsymbol{\delta'}
\end{array}\right](\boldsymbol{w}_{P_0}(P) \mid \tau)=
e^{i2\pi\boldsymbol{\tilde{\delta}}\cdot(\boldsymbol{w}_{P_0}(P)+\boldsymbol{\varepsilon'})+i\pi\boldsymbol{\tilde{\delta}}\cdot\tau\cdot\boldsymbol{\tilde{\delta}}}\Theta\left[\begin{array}{l}
\boldsymbol{\varepsilon} \\
\boldsymbol{\delta}
\end{array}\right](\boldsymbol{w}_{P_0}(P)+\boldsymbol{\tilde{\varepsilon}}+\tau\cdot\boldsymbol{\tilde{\delta}} \mid \tau). 
\end{equation}
This means that $P$ is no longer necessarily a Riemann zero for new characteristics.
\begin{figure}[t]
\centering 
\includegraphics[width=1\textwidth]{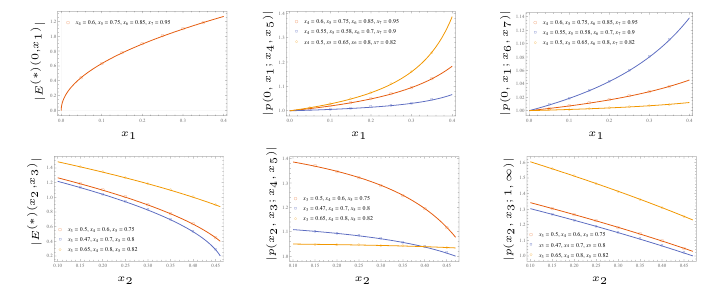}
\caption{\label{figB4} Plots for the conjectures in eq.\eqref{eq4.35}-\eqref{eq4.38} for $N=5$ and $n=2$. On the vertical axis, the solid lines are conjectured functions and the plotted points are numerically computed values of modified prime forms and cross-ratio functions. For the plots at the top $x_2=0.4$ and $x_3=0.45$ was used. For the plots at the bottom $x_1=0.1$, $x_6=0.85$ and $x_7=0.90$ was used.}
\end{figure}

This allows us to deal with Riemann zeros at branch points during numerical implementation by choosing the characteristics wisely. While implementing the cross-ratio one must make sure that the characteristics on all the theta functions are the same. We have shown plots for a few cases for the numerical conjectures in eq.\eqref{eq4.35}-\eqref{eq4.38} in Figure \ref{figB2}, Figure \ref{figB3}, and Figure \ref{figB4}. We see from these figures that conjectures have a good match with the numerical results.
\section{Multi-charged moments for massless Dirac fermion} \label{C}
In this appendix, we evaluate the multi-charged moments of $2$d massless Dirac fermions for multiple-disjoint intervals. The charged and multi-charged moments of $2$d massless Dirac fermions have been studied in the single interval \cite{murciano2020entanglement} and two disjoint intervals \cite{ares2022multi} cases. The extension to multi-interval is relatively straightforward, but for the sake of completeness, we present these calculations in detail.

Massless Dirac fermion in $2$d is a conformally invariant theory with central charge $c=1$. Dirac fermions, however, unlike compact boson, are not modular invariant. The theory is described by the action
\begin{equation}\label{eqC.1}
S=\int \mathrm{d}^2x \bar{\psi}\gamma^{\mu}\partial_{\mu}\psi.
\end{equation}
The gamma matrices are taken to be the Pauli matrices, $\gamma^0=\sigma_1$ and $\gamma^0=\sigma_2$. Dirac fermions posses a $U(1)$ symmetry under the transformation $\psi\to e^{i\alpha}\psi$ and $\bar{\psi}\to e^{-i\alpha}\bar{\psi}$.

To evaluate multi-charged moments for Dirac fermion we may consider n-copies of the field on the plane and introduce twist fields $\mathcal{T}_{\alpha}$ and $\bar{\mathcal{T}}_{-\alpha}$ \cite{casini2005entanglement, murciano2020entanglement}. The twist field act on the fields $\{\psi_i\}$ as
\begin{equation}\label{eqC.2}
\mathcal{T}_\alpha=\left[
\begin{array}{llll}
0 & e^{i\frac{\alpha}{n}} &  &\\
  & 0 & e^{i\frac{\alpha}{n}} &\\
  & &\ddots &\ddots\\
(-1)^{n}e^{i\frac{\alpha}{n}} & & &
\end{array}
\right].
\end{equation}
In this context $\mathcal{T}_{\alpha}$ is sometimes called the twist matrix. The multi-charge moments are given by the correlation function of twist fields placed on the boundary points of the intervals
\begin{equation}\label{eqC.3}
Z^{f}_{N,n}(\alpha_1,\cdots,\alpha_N)\propto\left\langle\prod_{i=1}^N\mathcal{T}_{\alpha_i}(u_i)\bar{\mathcal{T}}_{-\alpha_i}(v_i)\right\rangle.
\end{equation}
We may decouple the $n$-copies of the field $\psi$ by taking a unitary transformation that diagonalises the twist matrix. The eigenvalues of the twist matrix $\mathcal{T}_{\alpha}$ are $\lambda_{k,\alpha}=\frac{2\pi k}{n}+\frac{\alpha}{n}$, where $k\in\left(-\frac{n-1}{2},\cdots,\frac{n-1}{2}\right)$. The field $\psi_k$ in this basis satisfies the following monodromy conditions around the boundary points
\begin{align}
\psi_k\left((z-u_i)e^{i2\pi}\right)&= e^{i\left(\frac{2\pi k}{n}+\frac{\alpha_i}{n}\right)}\psi_k\left((z-u_i)\right),\label{eqC.4}\\
\psi_k\left((z-v_i)e^{i2\pi}\right)&= e^{-i\left(\frac{2\pi k}{n}+\frac{\alpha_i}{n}\right)}\psi_k\left((z-v_i)\right).\label{eqC.5}
\end{align}
The multi-charged moments in the new field basis are given by
\begin{equation} \label{eqC.5i}
Z^{f}_{N,n}(\alpha_1,\cdots,\alpha_N)\propto\prod_{k}\left\langle\prod_{i=1}^N\mathcal{T}_{k,\alpha_i}(u_i)\bar{\mathcal{T}}_{k,-\alpha_i}(v_i)\right\rangle ,
\end{equation}
where $\mathcal{T}_{k,\alpha_i}$ are the corresponding twist field which generates the monodromy condition in eq.\eqref{eqC.4}, and eq.\eqref{eqC.5}. The field $\psi_k$ may be written as $\psi_k(x)=e^{i\int_{x_0}^{x}\mathrm{d}\tilde{x}^{\mu}A^{k}_{\mu}(\tilde{x})}\tilde{\psi}_k(x)$ \cite{casini2005entanglement}. Here we have introduced single valued field $\tilde{\psi}_k(x)$ and the gauge fields $A^{k}_{\mu}$. It then follows from eq.\eqref{eqC.4}, and eq.\eqref{eqC.5} that the gauge field $A^{k}_{\mu}$ satisfies
\begin{align}
\oint_{u_i}\mathrm{d}\tilde{x}^{\mu}A^{k}_{\mu}(\tilde{x})&=\frac{2\pi k}{n}+\frac{\alpha_i}{n},\label{eqC.6}\\
\oint_{v_i}\mathrm{d}\tilde{x}^{\mu}A^{k}_{\mu}(\tilde{x})&=-\frac{2\pi k}{n}-\frac{\alpha_i}{n}.\label{eqC.7}
\end{align}
Using the Stoke's theorem, we deduce for eq.\eqref{eqC.6}, and eq.\eqref{eqC.7} that
\begin{equation}\label{eqC.8}
\epsilon^{\mu\nu}\partial_{\mu}A^{k}_{\nu}(z)=\sum_{i=1}^N\left(\frac{2\pi k}{n}+\frac{\alpha_i}{n}\right)\delta\left(z-u_i\right)-\left(\frac{2\pi k}{n}+\frac{\alpha_i}{n}\right)\delta\left(z-v_i\right).
\end{equation}
The multi-charged moments in eq.\eqref{eqC.5i} are now just the product of the partition functions of the gauged fields $\tilde{\psi}_k$. Multi-charged moments are now given by
\begin{equation}\label{eqC.9}
Z^{f}_{N,n}(\alpha_1,\cdots,\alpha_N)\propto\prod_{k}\langle e^{i\int\mathrm{d}^2x \bar{\tilde{\psi}}_{k}\gamma^{\mu}\tilde{\psi}_{k}A^{k}_{\mu}}\rangle.
\end{equation}
We may evaluate the partition function using the bosonisation of the fermions, $\bar{\tilde{\psi}}_{k}\gamma^{\mu}\tilde{\psi}_{k}\leftrightarrow -\frac{1}{2\pi}\epsilon^{\mu\nu}\partial_{\nu}\varphi_k$, where $\varphi_k$ is the compact boson at compactfication radius $R=1$. Using this duality the multi-charged moments may be written as the correlation function of the boson vertex operators $\mathcal{V}_{\alpha}$,
\begin{equation} \label{eqC.10}
Z^{f}_{N,n}(\alpha_1,\cdots,\alpha_N)\propto\prod_{k}\left\langle\prod_{i=1}^{N}\mathcal{V}_{-\left(\frac{k}{n}+\frac{\alpha_i}{2\pi n}\right)}(u_i)\mathcal{V}_{\left(\frac{k}{n}+\frac{\alpha_i}{2\pi n}\right)}(v_i)\right\rangle.
\end{equation}
The correlation function of the vertex operators on the complex plane has been extensively studied in the literature, see for example \cite{francesco2012conformal}. The neutrality condition, necessary for non-vanishing correlations, is satisfied by eq.\eqref{eqC.10}. The multi-charged moments are then given by
\begin{equation}\label{eqC.11}
\begin{split}
Z^{f}_{N,n}(\alpha_1,\cdots,\alpha_N)=&c_{N,n;\alpha_1,\cdots,\alpha_N}\prod_{k}\left(\prod_{i=1}^N\ell_i^{-2\left(\frac{k}{n}+\frac{\alpha_i}{2\pi n}\right)^2}\prod_{i<j=1}^{N}y_{ij}^{-2\left(\frac{k}{n}+\frac{\alpha_i}{2\pi n}\right)\left(\frac{k}{n}+\frac{\alpha_j}{2\pi n}\right)}\right)\\
&=c_{N,n;\alpha_1,\cdots,\alpha_N}Z^{f}_{N,n}\prod_{i=1}^N\ell_i^{-\frac{\alpha_i^2}{2\pi^2 n}}\prod_{i<j=1}^{N}y_{ij}^{-\frac{\alpha_i\alpha_j}{2\pi^2 n}},
\end{split}
\end{equation}
where $Z^{f}_{N,n}$ is just the reduced density matrix of $N$ disjoint interval for massless Dirac fermion. It is given by \cite{casini2005entanglement}
\begin{equation} \label{eqC.12}
Z^{f}_{N,n}=\prod_{i=1}^N\ell_i^{-\frac{1}{6}\left(n-\frac{1}{n}\right)}\prod_{i<j=1}^{N}y_{ij}^{-\frac{1}{6}\left(n-\frac{1}{n}\right)}.
\end{equation}
\section{Numerical Model} \label{D}
In this appendix, we discuss the tight-binding model. Tight-binding model is the lattice theory of massless Dirac fermions. This model is used to perform the numerical checks against the analytical results in Section \ref{section4} and \ref{section5}.

Tight-binding model is given by the Hamiltonian $H=-\sum_i \hat{c}^{\dagger}_{i+1}\hat{c}_{i}+\hat{c}^{\dagger}_{i}\hat{c}_{i+1}$. The fermionic operators $\hat{c}_i$ satisfies the commutation relations $\{\hat{c}_i,\hat{c}^{\dagger}_j\}=\delta_{i,j}$. This model has the correlation matrix $C_{ij}=\left\langle\hat{c}^{\dagger}_i\hat{c}_j \right\rangle$, given by
\begin{equation} \label{eqD.1}
C_{ij}=\frac{\sin{\left((i-j)\pi/2\right)}}{(i-j)\pi}.
\end{equation}
The moments of the reduced density matrix $\rho_A$ is given by \cite{peschel2003calculation, vidal2003entanglement}
\begin{equation} \label{eqD.2}
\mathrm{Tr}(\rho_A^n)=\prod_j\left[\left(\frac{1+\varepsilon_j}{2}\right)^n+\left(\frac{1-\varepsilon_j}{2}\right)^n\right],
\end{equation}
where $\left(\frac{1+\varepsilon_j}{2}\right)$ are the eigenvalues of the correlation matrix $C_{ij}$ restricted to subsystem $A$ (i.e. $i,j\in A$).

Tight-binding model possesses a global $U(1)$ symmetry. The corresponding conserved charge is given by $Q=\sum_i\hat{c}^{\dagger}_{i}\hat{c}_{i}-\frac{1}{2}$. The charge moments, similar to eq.\eqref{eqD.2}, are given by \citep{parez2021exact}
\begin{equation} \label{eqD.3}
\mathrm{Tr}(\rho_A^ne^{i\alpha\hat{Q}_A})=\prod_j\left[\left(\frac{1+\varepsilon_j}{2}\right)^{n}e^{i\frac{\alpha}{2}}+\left(\frac{1-\varepsilon_j}{2}\right)^ne^{-i\frac{\alpha}{2}}\right].
\end{equation}
To find the multi-charge moment $Z_{N,1}(\alpha_1,\cdots,\alpha_N)$ we first write the operator $e^{i\sum_j \alpha_j\hat{Q}_{A_j}}$, where $j\in\{1,2,\cdots,N\}$ as a Gaussian operator, with the corresponding correlation matrix $B$. The correlation matrix $B$ is given by
\begin{equation} \label{eqD.4}
B_{ij}=\delta_{i,j}\frac{e^{i\alpha_i}}{1+e^{i\alpha_i}}\qquad i\in A_i.
\end{equation}
This leads to, using the algebra rules for Gaussian operators \cite{fagotti2010entanglement}, the multi-charged moment
\begin{equation} \label{eqD.5}
\mathrm{Tr}\rho_{A}e^{\sum_{j=1}^N i\alpha_j\hat{Q}_{A_j}}=\left(\prod_{j=1}^{N}\left(e^{-i\frac{\alpha_j\ell_j}{2}}+e^{i\frac{\alpha_j\ell_j}{2}}\right)\right)\mathrm{det}\left(\frac{\mathbb{I}+W_{\{\alpha_i\}}}{2}\right).
\end{equation}
The matrix $W_{\{\alpha_i\}}$ is given by
\begin{equation} \label{eqD.6}
W_{\{\alpha_i\}}=
\left[
\begin{array}{llll}
G_{11} & G_{11} & \cdots  & G_{1N} \\
G_{21}  & G_{22} &  & \\
 \vdots & &\ddots &\\
G_{N1} & & &G_{NN}
\end{array}
\right]
\left[
\begin{array}{llll}
\frac{e^{i\alpha_1}-1}{e^{i\alpha_1}+1}\mathbb{I}_{\ell_1\times\ell_1} &  &  &\\
  & \frac{e^{i\alpha_2}-1}{e^{i\alpha_2}+1}\mathbb{I}_{\ell_2\times\ell_2} &  &\\
  & &\ddots &\\
 & & &\frac{e^{i\alpha_N}-1}{e^{i\alpha_N}+1}\mathbb{I}_{\ell_N\times\ell_N}
\end{array}
\right],
\end{equation}
where the block matrices $G_{IJ}=2C_{IJ}-\delta_{IJ}\mathbb{I}_{\ell_{I}\times\ell_{I}}$ and the index $I\in\{1,2,\cdots,N\}$. The matrix notation $C_{IJ}$ refers to the correlation matrix between the sites in $A_{I}$ with the sites in $A_{J}$. The eq.\eqref{eqD.5}-\eqref{eqD.6} for $N=3$ have been used to plot the multi-charge moments in Figure \ref{fig:iii}, charged moments in Figure \ref{fig:iv} and finally $\mathcal{Z}_{3,2}(q)$ in Figure \ref{fig:v}.
\acknowledgments
HG is supported by the Prime Minister’s Research Fellowship offered by the Ministry of
Education, Govt. of India.  UY's work is partly supported by Institute Chair Professorship of IIT Bombay.
\bibliographystyle{unsrt}
\bibliography{Biblio1.bib}




\end{document}